\documentclass{article_saj}
\usepackage{graphicx,saj,multicol,subeqnarray}
\usepackage{natbib}
\usepackage{float}
\usepackage{xcolor}
\usepackage{widetext}
\usepackage{url}
\usepackage{bm}
\usepackage{tikz} % for checkmark
\usepackage{pifont} % for xmark
\usepackage{amsfonts}
\usepackage{amssymb}
\usepackage{titlesec}
\titlelabel{\thetitle.\quad}
\definecolor{xlinkcolor}{cmyk}{1,0.6,0,0}
\usepackage[bookmarks=false,   % show bookmarks bar?
     pdfnewwindow=true,        % links in new window
     colorlinks=true,          % false: boxed links; true: colored links
     linkcolor=xlinkcolor,     % color of internal links
     citecolor=xlinkcolor,     % color of links to bibliography
     filecolor=xlinkcolor,     % color of file links
     urlcolor=xlinkcolor,      % color of external links
final=true
]{hyperref}

\def\papertype{\ \hfill\ scientific paper}

\def\udc{...}
\begin{document}
\parindent=.5cm
\baselineskip=3.8truemm
\columnsep=.5truecm
\newenvironment{lefteqnarray}{\arraycolsep=0pt\begin{eqnarray}}
{\end{eqnarray}\protect\aftergroup\ignorespaces}
\newenvironment{lefteqnarray*}{\arraycolsep=0pt\begin{eqnarray*}}
{\end{eqnarray*}\protect\aftergroup\ignorespaces}
\newenvironment{leftsubeqnarray}{\arraycolsep=0pt\begin{subeqnarray}}
{\end{subeqnarray}\protect\aftergroup\ignorespaces}
%

% Runningtitle

\markboth{\eightrm Symbiotic stars 4U1954+319 and ZZ~CMi} 
{\eightrm Zamanov {\lowercase{\eightit{et al.}}}}

\begin{strip}

{\ }

\vskip-1cm

\publ

\type

{\ }

% Title

\title{Luminosity class of the symbiotic stars 4U1954+319 and ZZ~CMi}

% Authors

\authors{R. Zamanov$^{1}$,                    % et al. } 
         K. A. Stoyanov$^{1}$, G. Latev$^{1}$, J. Marti$^2$,
         A. Takey$^{3}$,  E. G. Elhosseiny$^{3}$,  
         }
\authors{M. D. Christova$^4$, M. Minev$^{1}$, V. Vuj\v{c}i\'c$^5$,  
	 M. Moyseev$^{1}$, and V. Marchev$^{1}$}

\vskip3mm
% Address

\address{$^1$Institute of Astronomy and National Astronomical Observatory, Bulgarian Academy of Sciences, 72 Tsarigradsko Chaussee Blvd., 1784 Sofia, Bulgaria}
% E-mail
\Email{rkz@astro.bas.bg, kstoyanov@astro.bas.bg}

\address{$^2$Departamento de F\'isica, Escuela Polit\'ecnica Superior de Ja\'en, Universidad de Ja\'en, Campus Las Lagunillas s/n, A3-420, 23071, Ja\'en, Spain  }

\address{$^3$National Research Institute of Astronomy and Geophysics (NRIAG), 
              11421 Helwan, Cairo, Egypt}

\address{$^4$Department of Applied Physics, Technical University of Sofia, blvd. Kl. Ohridski 8, 
         1000 Sofia, Bulgaria}

\address{$^5$Astronomical Observatory, Volgina 7, 11060 Belgrade 38, Serbia}

% Received and Accepted dates

\dates{February 7, 2024}{.... .., 2024}

% Abstract

\summary{We performed optical photometry and spectral observations 
of the symbiotic stars  4U1954+319 and ZZ~CMi. 
For 4U1954+319 using high-resolution spectra we measure the 
equivalent widths of diffuse interstellar bands (DIBs)
and 
estimate the interstellar reddening $E(B-V)=0.83 \pm 0.09$.
Using $GAIA$ distances and our photometry, 
we find 
(1) absolute V band magnitude of 4U1954+319 $M_V=-5.23 \pm 0.08$ and 
that the mass donor is a supergiant of luminosity class Ib, and 
(2) for ZZ~CMi  $M_V= -0.27 \pm 0.2$ 
and that the mass donor 
is a giant of luminosity class III. 
}

% Keywords (see keywords.pdf file)


\end{strip}

\tenrm

% Sections

\section{INTRODUCTION}

Symbiotic stars are long-period interacting binary systems composed of a hot component, 
a cool giant and a nebula formed from material lost by the donor star 
and ionized by the radiation of the hot component 
(Miko{\l}ajewska 2012). 
The hot component can be a white dwarf, neutron star, or main-sequence star. 
In the majority, the donor star is a red giant 
but occasionally it could be an asymptotic giant branch star.
The spectrum of symbiotic stars is a combination of emission lines from 
the hot component and the nebula, 
and absorption lines from the donor star. 
The donor star loses mass 
through stellar wind or Roche-lobe overflow. 
The hot component accretes matter and produces high-energy emission  
and the symbiotic phenomenon (Miko{\l}ajewska 2007).  

To the current moment (January 2024)
SIMBAD Astronomical database (Wenger et al. 2000)  gives: 
for 4U1954+319 spectral type  M4/5III (Masetti et al. 2006),
and for  ZZ CMi spectral type  M6I-IIep (Shenavrin et al. 2011). 
We performed spectral observations and optical photometry, 
aiming to find the luminosity classes of these symbiotic stars.

\section{Observations}
Five optical spectra of 4U1954+319 and two of ZZ~CMi were secured with
the ESpeRo Echelle spectrograph (Bonev et al. 2017) on the
2.0-m telescope of the Rozhen National Astronomical Observatory, 
Bulgaria. The journal of observations is presented in Table~\ref{t.sp}, 
where are given the start of the observation, exposure time, signal-to-noise ratio
around 6570~\AA, the measured equivalent width of  $ H\alpha$ line 
and three DIBs.

BV photometry  was obtained with 
the 1.88-m telescope (Azzam et al. 2010) at the Kottamia Astronomical
Observatory, Egypt, with 
the 50/70-cm Schmid telescope at Rozhen National Astronomical Observatory, 
Bulgaria, and with 
the 0.4-m University of Ja\'en Telescope, Spain (Marti et al. 2017) 
A few stars from the APASS~DR10 located close to our targets
were used as comparison stars.
The values of the observed $B$ 
and $V$ band magnitudes are given in Table~\ref{t.ph}.
 
% $B=12.13 \pm 0.10$ and $V=10.01\pm0.01$
% $B=11.67 \pm 0.22$ and $V=10.31\pm0.16$ 

%----------------------------------------------------------
\begin{table*}
\caption{Spectral Observations of 4U1954+319. In the table are given the
date of observation (in format  YYYY-MM-DD~HH:MM), 
the exposure time in minutes, signal-to-noise ratio, 
and equivalent width of the  DIB5780, DIB5797, and DIB6613. }
\begin{tabular}{lcccccccc}
\\
 \hline
\\
object/date      & exposure  & S/N & EW$(H\alpha)$ & EW$_{5780}$ & EW$_{5797}$ & EW$_{6613}$  & \\
                 & [min]     &     & [\AA]         &   [\AA]     &	  [\AA]   &  [\AA]	 & \\
\\
{\bf 4U1954+319} &  \\
2015-08-03 23:28 &  40       & 42  & $+1.3$ &	0.31	 & 0.16        &   0.16       & \\
2016-06-18 23:41 &  40       & 70  & $+0.3$ &	0.39	 & 0.22        &   0.14       & \\
2018-04-03 00:06 &  60       & 60  & $+1.4$ &	0.35	 & 0.20        &   0.14       & \\ 
2018-09-01 20:45 &  60       & 75  & $+1.0$ &	0.28	 & 0.14        &   0.13       & \\
2022-04-12 22:55 &  60       & 53  & $+0.5$ &	0.31	 & 0.14        &   0.15       & \\
\\
{\bf ZZ~CMi} & \\
2023-12-27 00:34 &  45       & 70  & $-7.8$ &    ---      & ---	  &   ---	 & \\
2024-01-26 18:56 &  105      & 45  & $-4.8$ &    ---      & ---	  &   ---	 & \\
\hline
\\
\end{tabular} 
\label{t.sp}
%\end{table*}
%-----------------------------------------------------------
%----------------------------------------------------------
%\begin{table*}
\caption{BV photometry of 4U1954+319 and  ZZ~CMi. In the table are given 
the date of observation (in the format  YYYY-MM-DD~HH:MM), the telescope, number of the exposures, and
B and V band magnitudes with the corresponding errors.  }
\begin{tabular}{lcccccccc}
\\
  \hline
 object/date        & telescope & N & B  & V &  \\ 
                    &           &   &    &   &  \\
{\bf  4U1954+319} \\
 2023-08-08 18:47  &  1.88m  & 3 &  $12.07\pm0.03$ & $ 9.93\pm0.02$ & \\
 2023-12-05 16:32  &  1.88m  & 3 &  $12.08\pm0.01$ & $ 9.97\pm0.01$ & \\
 2023-12-06 16:32  &  1.88m  & 4 &  $12.09\pm0.01$ & $ 9.97\pm0.01$ & \\
 2024-01-22 18:21  &  40cm   & 2 &                 & $10.05\pm0.03$ & \\
 \\
{\bf  ZZ~CMi}   \\
 2023-12-12 23:19 & 50/70cm  & 3 &  $11.04\pm0.01$ & $ 9.57\pm0.01$ & \\
 2023-12-12 23:30 & 50/70cm  & 2 &  $11.06\pm0.01$ & $ 9.57\pm0.01$ & \\
 2024-01-22 18:50 & 40cm     & 2 &  $11.3 \pm 0.1$ & $10.06\pm0.05$ & \\
\\
\hline
\\
\end{tabular}
\label{t.ph}
\end{table*}
%-----------------------------------------------------------

\section{Interstellar extinction}

The NASA/IPAC  Galactic Reddening and Extinction Calculator
uses the Schlegel, Finkbeiner \& Davis (1998) Galactic reddening maps 
to determine the total Galactic line-of-sight extinction. 
It gives for 4U1954+319 
$E(B-V) \leq 1.68$ 
and for ZZ~CMi -- $E(B-V) \leq 0.04$. 

\begin{figure}
  \vspace{8.1cm}   
  \includegraphics{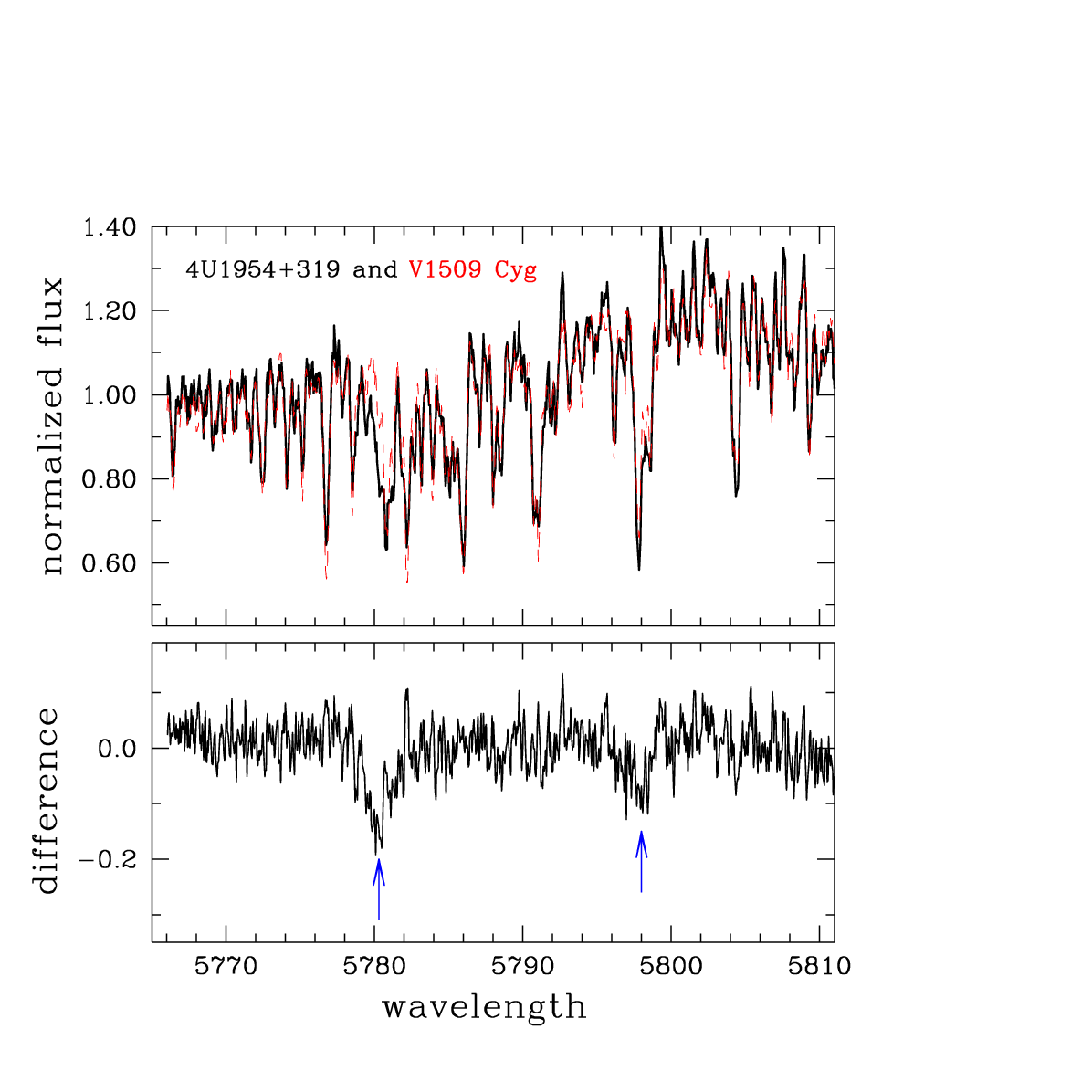} 
  \caption[]{The spectra of 4U1954+319 (black solid line) and V1509~Cyg (red dashed line). 
             The subtraction of the red giant spectrum
             reveals the presence of diffuse interstellar bands in the spectrum of 4U1954+319. 
  	     DIB~5780 and DIB~5797 are marked with blue arrows.   
           }      
\label{f.sp}
\end{figure}

From the spectra of 4U1954+319, we subtract the spectrum of a red giant 
(in this case we use a spectrum of V1509~Cyg obtained on 26 November 2023
with the same setup). The subtraction reveals the presence of
DIBs and gives us the possibility to measure their EWs. 
An example of the subtraction is shown in Fig.~\ref{f.sp}.
The measured EWs are given in Table~\ref{t.sp}.
The errors of EWs are 
$\pm 5\%$ for DIB~5780,  $\pm 7\%$ for DIB6613, and $\pm 10\%$ for DIB5797. 
As expected, on our spectra of ZZ~CMi we do not detect any DIBs. 

Puspitarini, Lallement \& Chen (2013) found a set of relations between 
the equivalent width of the DIBs and the interstellar reddening, among them: \\
$E(B-V) = 2.3 \; EW_{5780} + 0.0086$, \\
$E(B-V) = 5.1 \; EW_{6613} + 0.0008$, \\
$E(B-V) = 6.3 \; EW_{5797} + 0.0203$. \\
Using these relations and 
our measurements of the EWs (given in Table~\ref{t.sp}) as well as 
taking into account the individual errors 
we estimate the interstellar extinction 
toward 4U1954+319 $E(B-V)=0.83 \pm 0.09$.

Hereafter we will  use  $E(B-V)=0.83$ for 4U1954+319, 
and $E(B-V)=0$ for ZZ~CMi.

%---------------------------------------------------------
%---------------------------------------------------------

\section{Absolute V band magnitude}

\begin{figure*}  
\vspace{10.4cm}   
\includegraphics{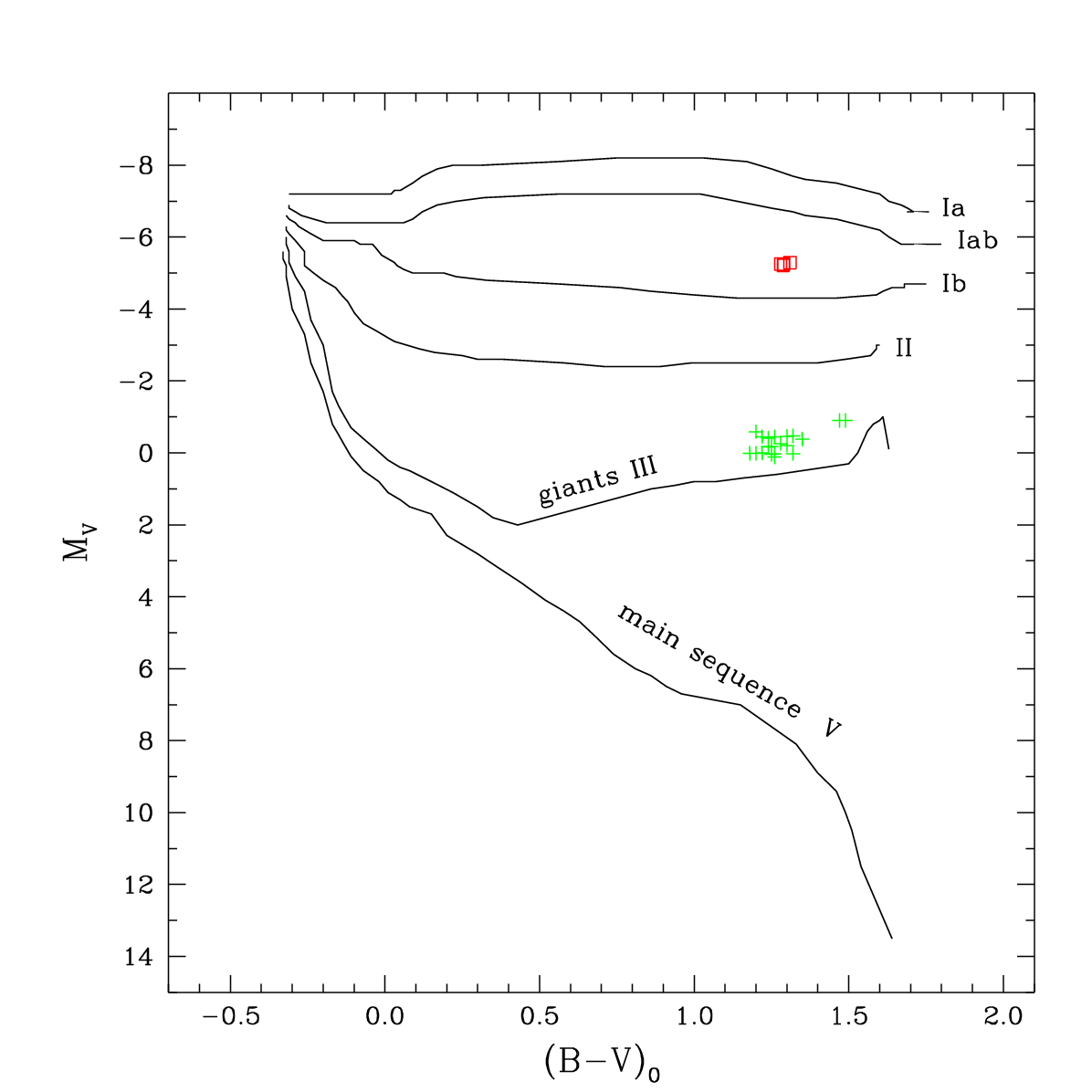}  
\caption{Colour-magnitude diagram, $M_V$ versus $(B-V)_0$,  
for 4U1954+319 (red squares)
and ZZ~CMi (green plusses). 
The lines for the different luminosity classes
are also plotted. }
\label{f.cm}
\end{figure*}

Using the GAIA eDR3 (Gaia Collaboration et al. 2021)  % \citep{2021A&A...649A...1G}  
the model by Bailer-Jones et al. (2021)              % \citet{2021AJ....161..147B}  
provides  distances
$d=3390 \pm 60$~pc  to  4U1954+319 and 
$d=1240 \pm 20$~pc  to  ZZ~CMi. 
To calculate the absolute V band magnitude we use the 
standard formula, $M_V=m_v-A_V-5 \log(d/10)$, where $A_V$ is the extinction in V band,  
$A_V=3.1 E(B-V)$.
In Fig.~\ref{f.cm} is plotted the colour-magnitude diagram.
The values of  $M_V$ and $(B-V)_0$ 
for the different luminosity classes
are from Straizys \& Kuriliene (1981) and  from 
Schmidt-Kaler (1982), respectively. 

For  4U1954+319 using our data (Table~\ref{t.ph}) 
and APASS DR10 photometry (V=10.01, B=12.13),
we obtain  $M_V=-5.23 \pm 0.08$. The position of the object 
in Fig.~\ref{f.cm} indicates that it is slightly above 
luminosity class Ib (supergiants) and  below Iab (moderate supergiants). 

For ZZ~CMi using the new observations from Table~\ref{t.ph}
as well as published data in Zamanov et al. (2021), 
we find  $M_V= -0.27 \pm 0.2$ . 
The position of the object 
on Fig.~\ref{f.cm} indicates that it is slightly above 
luminosity class III (giants) and  below luminosity class II (bright giants). 

In the symbiotic stars, the flux in V band comes from three sources -- 
cool giant, hot component, and nebula. 
In the two symbiotic stars discussed here, the main source in V band is the mass donor.
The $EW(H\alpha)$ given in Table~\ref{t.sp} indicates 
that  the contribution of the other sources is not more than 10\% for 4U1954+319
and not more than 15\% for ZZ~CMi.  
We conclude that the luminosity class of the mass donor in 4U1954+319
is Ib, and in ZZ~CMi is III. 

%---------------------------------------------------------
%---------------------------------------------------------

\section{Discussion}

4U~1954+319 is a member of a small group of binary systems with accreting neutron stars and late-type
giant stars (Masetti et al. 2006). They are called symbiotic X-ray binaries. 
Only about a dozen of these systems are known (Yungelson et al. 2019). 
Hinkle et al. (2020) classified 4U~1954+319 as a symbiotic X-ray binary that consist of an M supergiant. 
As per our knowledge, apart from 4U~1954+319 there 
is only one more X-ray binary that harbors a late-type supergiant as a mass donor, 
namely CXOGC~J174528.7-290942 (Gottlieb et al. 2020).
The mass of the M supergiant is considered to be $\sim 9$~M$_\odot$  (Hinkle et al. 2020). 
Although a rare class system, 4U~1954+319 follows the evolution scenario 
of the High-mass X-ray binaries
(Tauris et al. 2017) from two massive
and hot progenitors but it is more evolved from 
the typical High-mass X-ray binaries. 
The orbital period of 4U~1954+319 is not yet confirmed 
but Hinkle et al. (2020) estimate it to be $\gtrsim$ 3~yr. \\

ZZ~CMi is a symbiotic star that consists of a white dwarf and a red giant star, with 
the orbital period likely to be $\sim$ 440~d (Wiecek et al. 2010).
Tshernova (1949) found that ZZ~CMi is a long-period variable star. 
The GCVS catalog give a classification of ZZ~CMi M6I-IIep (Kholopov et al. 1998, Samus et al. 2017). 
Taranova \& Shenavrin (2001) found M4.5-5~III. Later, Shenavrin et al. (2011) obtained M6I-IIep.

%Previous classification of ZZ~CMi as M6I-Iiep by Shenavrin and Taranova (eg. Taranova and Shenavrin, 2001, %Astronomy Letters, 27, 5) after GCVS Catalog (Kholopov et al., 1985 – 1990). GCVS catalog refers to the %paper by Chernova T. S., Perem. Zvezdy 7, N3, 140, 1949.

The optical spectra of ZZ~CMi are dominated by the red giant 
with weak and variable emission lines of 
$H\alpha$, $H\beta$, [OIII], and  [Ne III] (Iijima 1984). 
Zamanov et al. (2021) noted that an outflow with 
velocity of about  150 km~s$^{-1}$ 
and  U band flickering with an amplitude of about $0.1$~mag are visible sometimes. 
ZZ~CMi is an X-ray source from the $ \beta / \delta$-type, 
which means that there are two X-ray thermal components, soft
and hard  (Luna et al. 2013).  
Some peculiarities deviate ZZ~CMi from the classical symbiotic stars 
-- the colours are bluer at minimum which is not typical for symbiotics 
and the strengths of the emission lines are unusual with 
$H\gamma >  H\beta$ (Belczy{\'n}ski et al. 2000).

\section{Conclusions}
On the basis of our spectral and photometric observations, we find 
for 4U1954+319 interstellar extinction $E(B-V)=0.83 \pm 0.09$, $M_V=-5.23 \pm 0.08$,  
and luminosity class Ib (supergiant) for the mass donor.
For  ZZ~CMi we find  $M_V= -0.27 \pm 0.2$ 
and mass donor of luminosity class III (giant). 

% Acknowledgements

\acknowledgements{This work has been partially supported 
by the Ministry of Education and Science of Bulgaria 
(Bulgarian National Roadmap for Research Infrastructure) 
and the Spanish Ministerio de Ciencia e Innovaci\'on,
Agencia Estatal de Investigac\'on (Ref. PID2022-136828NB-C42).
JM acknowledges Plan Andaluz de Investigaci\'on, Desarrollo e Innovaci\'on as research
group FQM-322, and FEDER funds. 
AT and EE acknowledge financial support from the Egyptian Science, Technology \& Innovation Funding Authority (STDF) under grant 45779.
The NASA/IPAC Extragalactic Database (NED) is funded 
by the National Aeronautics and Space Administration and operated by the California Institute of Technology.
This paper makes use of data from the AAVSO Photometric All Sky Survey (APASS), 
whose funding has been provided by the Robert Martin Ayers Sciences Fund and from the NSF (AST-1412587).} 

$ \; $

% References  \vskip2mm
%\newcommand\eprint{in press }   \bibsep=0pt
%\bibliographystyle{aa_url_saj}
%{\small
%\bibliography{sample_saj} }

\begin{thebibliography}{}

\bibitem[Azzam et al.(2010)]{2010ASSP...20..175A} Azzam, Y.~A., Ali, G.~B., Ismail, H.~A., 
Haroon, A.,  Selim, I.\ 2010, Astrophysics and Space Science Proceedings, 20, 175 

\bibitem[Bailer-Jones et al.(2021)]{2021AJ....161..147B} Bailer-Jones, C.~A.~L., Rybizki, J., 
Fouesneau, M., et al.\ 2021, \aj, 161, 147

\bibitem[Belczy{\'n}ski et al.(2000)]{2000A&AS..146..407B} Belczy{\'n}ski, K., Miko{\l}ajewska, J., Munari, U., et al.\ 2000, \aaps, 146, 407

\bibitem[Bonev et al.(2017)]{2017BlgAJ..26...67B} Bonev, T., Markov, H., Tomov, T., et al.\ 2017, Bulgarian Astronomical Journal, 26, 67

\bibitem[Gaia Collaboration et al.(2021)]{2021A&A...649A...1G} Gaia Collaboration, Brown, A.~G.~A., Vallenari, A., et al.\ 2021, \aap, 649, A1

\bibitem[Gottlieb et al.(2020)]{2020ApJ...896...32G} Gottlieb, A.~M., Eikenberry, S.~S., Ackley, K., et al.\ 2020, \apj, 896, 32

\bibitem[Hinkle et al.(2020)]{2020ApJ...904..143H} Hinkle, K.~H., Lebzelter, T., Fekel, F.~C., et al.\ 2020, \apj, 904, 143

\bibitem[Iijima(1984)]{1984IBVS.2491....1I} Iijima, T.\ 1984, Information Bulletin on Variable Stars, 2491, 1

\bibitem[Kholopov et al.(1998)]{1998GCVS4.C......0K} Kholopov, P.~N., Samus, N.~N., Frolov, M.~S., et al.\ 1998, Combined General Catalogue of Variable Stars, Moscow: Nauka Publishing House (1985-1988),
VizieR On-line Data Catalog: II/214A


\bibitem[Luna et al.(2013)]{2013A&A...559A...6L} Luna, G.~J.~M., Sokoloski, J.~L., Mukai, K., et al.\ 2013, \aap, 559, A6

\bibitem[Mart{\'\i} et al.(2017)]{2017BlgAJ..26...91M} Mart{\'\i}, J., Luque-Escamilla, P.~L., \& Garc{\'\i}a-Hern{\'a}ndez, M.~T.\ 2017, Bulgarian Astronomical Journal, 26, 91

\bibitem[Masetti et al.(2006)]{2006A&A...453..295M} Masetti, N., Orlandini, M., Palazzi, E., et al.\ 2006, \aap, 453, 295

\bibitem[Miko{\l}ajewska(2007)]{2007BaltA..16....1M} Miko{\l}ajewska, J.\ 2007, Baltic Astronomy, 16, 1

\bibitem[Miko{\l}ajewska(2012)]{2012BaltA..21....5M} Miko{\l}ajewska, J.\ 2012, Baltic Astronomy, 21, 5 

\bibitem[Puspitarini et al.(2013)]{2013A&A...555A..25P} Puspitarini, L., Lallement, R., \& Chen, H.-C.\ 2013, \aap, 555, A25

\bibitem[Samus et al.(2017)]{2017ARep...61...80S} Samus, N.~N., Kazarovets, E.~V., Durlevich, O.~V., et al.\ 2017, Astronomy Reports, 61, 80

\bibitem[Schlegel et al.(1998)]{1998ApJ...500..525S} Schlegel, D.~J., Finkbeiner, D.~P., \& Davis, M.\ 1998, \apj, 500, 525

\bibitem[Schmidt-Kaler(1982)]{1982......xxx..xxxS} Schmidt-Kaler, T. 1982, 
Landolt-B\"ornstein 2013 Group VI Astronomy and
Astrophysics, Volume 2B, Stars and Star Clusters (Berlin: Springer-Verlag), 1

\bibitem[Shenavrin et al.(2011)]{2011ARep...55...31S} Shenavrin, V.~I., Taranova, O.~G., \& Nadzhip, A.~E.\ 2011, Astronomy Reports, 55, 31

\bibitem[Straizys \& Kuriliene(1981)]{1981Ap&SS..80..353S} Straizys, V. \& Kuriliene, G.\ 1981, \apss, 80, 353

\bibitem[Taranova \& Shenavrin(2001)]{2001AstL...27..338T} Taranova, O.~G. \& Shenavrin, V.~I.\ 2001, Astronomy Letters, 27, 338

\bibitem[Tauris et al.(2017)]{2017ApJ...846..170T} Tauris, T.~M., Kramer, M., Freire, P.~C.~C., et al.\ 2017, \apj, 846, 170

\bibitem[Tshernova(1949)]{1949PZ......7..140T} Tshernova, T.~C.\ 1949, Peremennye Zvezdy, 7, 140

\bibitem[Wenger et al.(2000)]{2000A&AS..143....9W} Wenger, M., Ochsenbein, F., Egret, D., et al.\ 2000, \aaps, 143, 9

\bibitem[Wiecek et al.(2010)]{2010arXiv1003.0608W} Wiecek, M., Mikolajewski, M., Tomov, T., et al.\ 2010, arXiv:1003.0608

\bibitem[Yungelson et al.(2019)]{2019MNRAS.485..851Y} Yungelson, L.~R., Kuranov, A.~G., \& Postnov, K.~A.\ 2019, \mnras, 485, 851

\bibitem[Zamanov et al.(2021)]{2021AN....342..952Z} Zamanov, R.~K., Stoyanov, K.~A., Kostov, A., et al.\ 2021, Astronomische Nachrichten, 342, 952


\end{thebibliography}
% \begin{strip}
% \end{strip}

% \clearpage 

\bibsep=1pt

\end{document}